# Bibliometric Index for Academic Leadership

Yang Liu[1*], Fengrong Ou[1], Yan Deng[1], Bo Wu[2], Ruxi Liu[3], Hui Hua[1], Yuyuan Guan[1], Rentong Chen[1], Lars Gjesteby[4], Jiansheng Yang[5], Michael Vannier[6], Ge Wang[4*]

[1]Department of Environmental Health, China Medical University, Shenyang, Liaoning 110122, China

[2]Department of Anal and Rectal Diseases, First Hospital, China Medical University, Shenyang, Liaoning 110001, China

[3]Department of Immunology and Rheumatology, First Hospital, China Medical University, Shenyang, Liaoning 110001, China

[4]Rensselaer Polytechnic Institute, 110 8th Street, Troy, New York 12180, USA

[5]Peking University, Beijing 100871, China

[6]University of Chicago, Chicago, Illinois 60637, USA

*Correspondence to: Yang Liu (yangliu@cmu.edu.cn) and Ge Wang (ge-wang@ieee.org)

**Abstract**: Academic leadership is essential for research innovation and impact. Until now, there has been no dedicated measure of leadership by bibliometrics. Popular bibliometric indices are mainly based on academic output, such as the journal impact factor and the number of citations. Here we develop an academic leadership index based on readily available bibliometric data that is sensitive to not only academic output but also research efficiency. Our leadership index was tested in two studies on peer-reviewed journal papers by extramurally-funded principal investigators in the field of life sciences from China and the USA, respectively. The leadership performance of these principal investigators was quantified and compared relative to university rank and other factors. As a validation measure, we show that the highest average leadership index was achieved by principal investigators at top national universities in both countries. More interestingly, our results also indicate that on an individual basis, strong leadership and high efficiency are not necessarily associated with those at top-tier universities nor with the most funding. This leadership index may become the basis of a comprehensive merit system, facilitating academic evaluation and resource management.

**Introduction**

Today's progress in science, technology, and medicine is largely made via multi-/inter-/trans-disciplinary collaboration that greatly depends on teamwork and leadership. Therefore, optimization of research and development efforts is highly desirable to use resources wisely and maximize researchers' synergistic potential, which are in most cases the responsibilities of academic leaders at different levels. In most developed and developing countries, the principal investigators or equivalent roles direct research teams on numerous projects. The main topic of this paper is the bibliometrical measurement of leadership of these principal investigators (1-5).

Despite on-going debates over its validity and usage, bibliometrics has been widely used in almost all academic fields worldwide. Although peers' perception and content analysis are more relevant, citation analysis is the dominating approach because of its objectivity, accessibility, and popularity (6-28). Many universities and funding agencies use bibliometric indices to help gauge the importance of a journal, the impact of a paper, the caliber of a researcher, and the fitness of a candidate. A good example is the wide-spread applications of the h-index, which has been a hot research and public topic.

Citation analysis has become increasingly more powerful with the development of mobile internet, high-performance computing, and big data technologies. When the Science Citation Index (SCI) came to public use in 1961, data mining was manually done in a time-consuming fashion. Sophisticated yet efficient algorithms are now commonplace. The *CiteSeer* algorithm is the first success in automatic citation analysis (29). The *PageRank* algorithm is the key to Google's prosperity (30, 31), which is based on citation analysis. Looking further into bibliometric data, researchers even extracted social relationships between different authors and schools (32).

The central question here is how to bibliometrically measure the attributes of an academic leader? While the bibliometric output of science, engineering, and medicine is commonly measured (8, 12), leadership has not been rigorously quantified, as evidenced by the fact that there is no leadership index available in the bibliometric field. A reasonable/plausible bibliometric measurement of academic leadership should integrate both academic output and research efficiency, improving upon conventional indices that mostly underline academic output alone or rudimentary metrics that roughly estimate academic output per funding investment or group size. Our proposed leadership index may have important and immediate applications in the following scenarios.

First, academic leaders can be better evaluated with leadership-specific measures. It is well known that a control system needs a feedback loop. Without a quantitative measure, many questions cannot be answered. For example, assuming other conditions comparable, if leaders A and B respectively direct his/her groups of different sizes, and publish various papers in multiple journals of low and high impact factors, then who has stronger leadership from a societal perspective? Just as bibliometric indices are valuable in evaluating researchers, specific measures are desirable to quantify their leaders. These leadership measures would facilitate identification of top performers, and also help all leaders improve their skills.

Then, research resources should be distributed according to the capabilities of academic leaders so that for a given amount of resources the overall academic output can be maximized. Would it be reasonable to impose an upper limit of the total funding to a research group? If so, how would this upper limit be determined? If his/her group is over-funded, the increment of its scientific output will be at a lower rate while other more productive groups are underfunded, which means a waste of precious resources. A major task of a governmental funding agency is to optimally distribute research money. The funding portfolio ought to be adaptively refined with quantitative insight. All of these points are closely related to the leadership index we will define below.

The bibliometric quantification of academic leadership becomes increasingly desirable. A main engine of modern society is widely recognized as original innovations, which are now highly interdisciplinary and mostly performed by teams or groups. Both developed and developing countries invest a substantial portion of GDP to support basic and applied research and development. Just as the existing bibliometric indices have had a major impact on management of research institutes and universities, the bibliometric index for academic leadership will facilitate optimization of scientific research endeavors.

In this paper, we formulate, justify, and demonstrate how to use a bibliometric index for leadership. To our best knowledge, this leadership index is the first of its kind, adding a new dimension to bibliometric analysis. In the methodology section, we describe the leadership index briefly, with detailed justification, formulation and discussion in the supplementary materials section. In the results section, based on the datasets collected on some representative Chinese and American principle investigators and according to the experimental designs that were detailed in the supplementary materials section, we present statistical relationships between academic leadership, resources, and other factors at universities in different countries. In the last section, we discuss key observations and relevant issues.

**Leadership Index**

As fully elaborated in the methods section, here we propose a leadership index that balances academic output and research efficiency, each of which is bibliometrically measured. These two measures are then integrated into their geometric mean as a single figure of merit to bibliometrically quantify the leadership. As an initial example consistent with the common practice, the academic output is computed as the sum of journal impact factors (IF). To measure research efficiency bibliometrically, we introduce equivalent time T to reflect the human resource devoted to the academic output. Suppose that the total output is measured by the sum of IFs for a team of interest, and the team leader has published N papers as the corresponding author with $IF_i$, i=1, 2, …, N. We have the academic output O:

$$O = \sum_{i=1}^{N} IF_i \tag{1}$$

The leader may have not spent his/her time equally on these papers. It would be reasonable to claim that he/she has spent more time on papers with higher IFs. Explicitly, his/her relative effort on the jth paper with $IF_j$ is estimated as $IF_j / \sum_{i=1}^{N} IF_i$. With our previously-published axiomatic index (A-index) for coauthors' individual credits *(22)*, we can compute the leader's credit share of the jth paper, which is equal to $A_j * IF_j$. The rest of the work has been performed by his/her team members. In other words, the amount of the leader's equivalent time T spent on all his/her papers is

$$T = \frac{1}{\sum_{i=1}^{N} IF_i} \sum_{j=1}^{N} \frac{IF_j}{A_j} \tag{2}$$

As a result, we have the leader's research efficiency E:

$$E = \frac{O}{T} \tag{3}$$

The so-defined equivalent time is in the unit of the leader's time scale, and reflects the talent pool the leader directed. The outputs of different leaders are measured in the same unit of academic output, which is the sum of IFs. Hence, leaders or team qualities will be fairly differentiated in terms of IFs, the expected numbers of citations, while the equivalent time is measured on his/her own time scale. Roughly speaking, for a given academic output, a stronger leader has a higher research efficiency and spends less time than a weaker leader, which can be quantified by our leadership index L:

$$L = \sqrt{OE} = \frac{\left(\sum_{i=1}^{N} IF_i\right)^{1\frac{1}{2}}}{\sqrt{\sum_{j=1}^{N} \frac{IF_j}{A_j}}} \tag{4}$$

To further emphasize the importance of high-impact journal papers, IF factors can be individually weighted by our toughness index that is explained in the supplementary materials section.

**Results**

*Leadership of Top Ranked Universities*

Table 1 is a key result that compares leadership index values keyed to university ranking classes and two selected major countries. In an average sense, Table 1 shows the validity of our proposed leadership index. There are multiple interesting features revealed in Table 1. First, in terms of the leadership index L as well as the output and efficiency variables O and E, the first class outperformed its peers in both China and the USA with a strong statistical significance, while there was no significant difference in the group's effort as measured by the equivalent time T. Furthermore, L, O and E decreased from the first to third class in the USA, while these measures were comparable between the second and third classes in China. In terms of the leadership index, the first class in China was comparable or inferior to the third class in the USA. However, in terms of the scientific output without the use of the toughness index, the first class of the Chinese principal investigators outperformed their American peers (not statistically significant). It is underlined that the leadership index L is more comprehensive, integrating the scientific output O and the research efficiency E with the toughness index included to emphasize the potentially greater impact and novelty of high impact journals.

*Leadership versus Equivalent Time and Total Funding*

We analyzed the relationship between the academic leadership L and the involved research time T with our two datasets, as shown in Figure 1. For convenience, we discretized T with the sampling step 0.5, and real T values are clustered to the nearest neighbor. Note that there is only one sample in the Chinese dataset in each of the cases of T>7, and also one sample for T=7, 8.5, or 9 in the NIH data. Also, three samples of larger T values were not included in the analysis, which are (T, L) = (36, 4.745829), (50, 0.9015656), and (84.5, 1.838076). It is observed that the leadership index reached the peak for T=~5 in China, while the leadership index monotonically increased with the equivalent research time in the USA.

Interestingly, the peak in terms of the leadership index suggests a maximum group size, beyond which the leadership measure is compromised despite the fact that the output may still increase, but the efficiency will decrease past a critical point. Assuming that the corresponding author made the most important conceptual contribution to the paper, T=5 corresponds to the principal investigator's individual A-index value being 0.20, which in turn means that the maximum group size an average group leader could handle is 17 (see the first A-index table in the supplemental information, SI17-1) (22). If the assumption is that the principal investigator and the first author made equal contributions, this maximum group size is significantly reduced to 11 (see the second A-index table in the supplemental information, SI17-2) (22).

Then, we analyzed the relationship between the leadership and the funding total based on the collected NIH data, as shown in Figure 2. By nature of the data collection protocol, there is no statistical difference between the first two classes of the involved institutes in terms of the total funding received by the principal investigator. Generally speaking, the leadership index tends to take larger values for stronger funding support. The statistical correlation data show that the overall correlation between the leadership index and funding support was 0.188**. The correlation coefficients were 0.228** and 0.352** for the second and third classes of institutes, respectively. For the principal investigators in the first class of institutes, the correlation between the leadership index and funding was not significant. Hence, independent of funding, the highest average of leadership index values by NIH principal investigators in the first class indicates the overall best quality of the top tier university faculty members.

*Leadership versus Gender, Age, and Rank*

Tables 2 and 3 offer data on the variation of the Chinese academic leadership with respect to gender, age, and rank. In the data we have analyzed, the leadership measure was higher for men than women, and the age group of 55-60 years old demonstrated the strongest leadership, but interestingly the age group of >60 years old showed a performance similar to that of the age groups of 45-50 and 51-55 years old. Furthermore, as expected, the full professor had the strongest leadership among all ranks.

*Leadership Trends in China and the USA*

Figure 3 demonstrates the annual trends of academic leadership, scientific output, research efficiency, and equivalent time for Chinese and American principal investigators ranging from 2008 to 2015. It can be observed in Figure 3 that the Chinese leadership index has clearly increased over the period under investigation, while the American leadership is multiple folds stronger, but decreased slightly, perhaps due to the compromised funding situation. Generally speaking, China was on an upward trend in terms of academic leadership, scientific output, and research efficiency, while the USA was in fluctuation and on average went down. In contrast to these measures, equivalent time has been very stable for the American principal investigators.

**Discussion**

The validity of the proposed leadership index has been demonstrated in the above-reported bibliometric data keyed to the respective classes of institutional ranks in China and the USA. The principal investigators in the commonly-recognized top universities indeed demonstrated the strongest academic leadership that has now been bibliometrically quantified. The superior academic output and research efficiency of the principal investigators in top universities seems to be a common phenomenon, given the same resources in terms of equivalent time or total funding. This must be due to their academic talent and leadership, excellent local colleagues and students, and outstanding overall infrastructure.

Now, it seems natural to include the leadership index as a novel criterion balancing academic output and research efficiency for the ranking of universities, evaluation of principal investigators, and management of research resources (1-5). Importantly, our retrospective data analysis indicates that some Chinese groups might have used excessive human resources, as indicated by the peaked curves of the leadership index versus either equivalent time or number of grants (led to reduced efficiency; see Figure 1). Hence, our proposed leadership index could be a tool for an optimized management of research activities and funding decisions. On the other hand, both the leadership index and the academic output increased with resources (either equivalent time or total funding) in the USA (Figure 1). This positive correlation shows that the potential of the NIH research work force was not saturated by NIH funding, which had indeed been limited over the period of time under investigation.

It is very interesting to compare Chinese and American leadership changes over time in the life sciences field at large. Figure 3 shows that the Chinese leadership increased over time at a moderate rate, while the American leadership fluctuated and decreased. The gap between the Chinese and American curves was large in the common period of time. It is emphasized that the Chinese and American funds are in different currencies and systems, and a direct comparison is therefore difficult: on one hand, US dollars are more valuable, and the NIH support was stronger and can be renewed; on the other hand, human resources were cheaper in China than those in the USA, and the NSFC support did not allow renewal. Nevertheless, the equivalent time concept we have introduced above makes the comparison of research efficiency quite fair, as does the sum of weighted IFs which is the same for all the datasets in both countries. Hence, we believe that the trends do reflect the truth to a good degree.

It is underlined that our leadership index is an integrated measure more insightful than the popular narrow-focused indices such as the number of citations (7, 22). We have specifically introduced the toughness index to distinguish the essential difference between multiple low-impact papers and a single high-impact or high-novelty contribution. The incorporation of the toughness index into the leadership index is a key ingredient to emphasize the original research of high potential. A merit of our leadership index is that it can be dynamically measured in terms of current bibliometric data, and statistically analyzed in reference to other information readily available on the internet. Hence, our proposed figures of merit have major practical implications.

In this work, the scientific output was measured as the sum of journal paper values, which was in turn measured as weighted IFs. We believe that the weighted IF measurement is better than either the number of papers or the sum of simple IFs, because not all the papers are the same, and a high impact paper cannot be equivalent to an accumulation of low impact papers. Nevertheless, IF or weighted IF values are expected numbers. While some papers in Nature or Science were associated with Nobel prizes, some papers in the same journals are not even reproducible or fundamentally flawed. Hence, weighted citation-based measures should be more reliable than weighted IF-based counterparts. A good number of Nobel prizes were given to results reported in low impact journals, but those results were typically highly cited. An extensive citation record over a long time may suggest strong leadership. Technically speaking, citation-based measurement is more challenging than IF-based alternatives, since the total number of citations is a function of time. The longer the time, the better citation data but the greater the evaluation delay that may defeat the purpose. With the rapid development of internet, cloud computing, and machine learning technologies, better data collection and analysis methods may help integrate IF and citation data to improve the predictive power of leadership quantification.

Finally, our leadership index has a number of limitations similar to other bibliometric indices. The leadership index is the bibliometric aspect of leadership, and should not be interpreted the same as traditional leadership qualities (1-5). There are many good cases in which great work including a number of intellectual milestones did not show well in terms of bibliometric analysis. Hence, no matter how sophisticated bibliometric measures are, the real academic impact and practical value are the ultimate criteria, and peers' perception ought to be the gold standard of the outstanding leadership. Nevertheless, the bibliometric measures will significantly aid the formation of peers' perception, and serve as important tools for management as a science.

It is well known that bibliometric indices have quite different ranges of values with respect to multiple factors such as the popularity of the academic field (25). Hence, the leadership index should be presented with respect to appropriate baselines or after field-specific normalization. Our simplification to the equivalent time spent by a team has been shown to be a good approximation with interesting results, but this aspect can be further refined. For example, to compute the size of a team led by a principal investigator we could check if co-authors of different papers are the same or different, and whether these co-authors are research faculty, postdoc fellows, or graduate students. We believe that the leadership index we have established is a starting point that should be developed into more advanced variants and applied in various applications.

**Materials and Methods**

*Leadership Index*

As mentioned earlier, we postulate that the leadership index should balance two important aspects: (1) academic output and (2) research efficiency. This combination is consistent with both the business model (total revenue versus capital return) and the minimum energy principle (desirable outcome at minimum cost). That is, we want to maximize the output while minimizing the expenditure. The academic output from a team must be high for it to be influential. At the same time, the research efficiency of the team must be also taken into account, and if the team works inefficiently it will be invalid to claim that their leader is great. The first aspect was well studied in the bibliometric field, and is measured with a number of bibliometric indices. However, the second aspect has not yet been bibliometrically addressed.

The first aspect is typically measured in terms of publications. More specifically, the academic output, or bibliometric impact, of a team is generally perceived to be higher if they have a higher number of papers, a higher amount of prestigious journals in which these papers are published, and a higher number of citations for these papers. The number of papers is the simplest metric to count but the roughest to value. The number of citations is commonly respected, but it takes time for citations to be accumulated. The journal impact factor (IF) is a good indicator of a paper, but it is less relevant than the number of citations the paper could really attract. Let $O$ denote an output index, which is in terms of an appropriate bibliometric value $V$, such as the count of papers, the sum of IFs, or the number of citations, which are applicable in different contexts. If only one team contributes to papers, we have

$$O = \sum_{i=1}^{N} V_i \qquad (1)$$

where $N$ is the number of papers from the team, and $V$, as an example, can be journal impact factor (IF), the weighted IF (such as with the toughness index as the weighting factor to be explained below), the number of citations, and so on.

Our earlier work on a coauthor's individual credit in a paper is measured by the $A$ index. In the context of this paper, the emphasis is on leadership and the credit of his/her team. Hence, we focus on the order of the affiliations associated with the coauthors of the paper, and credit the team leader's affiliation (that is, his/her team) using the $A$ index. In this way, the output index $O$ of the team for a given period of time can be computed as follows:

$$O = \sum_{i=1}^{N} A_i V_i \qquad (2)$$

where $A_i$ is the $A$ index of the $i$th paper for the team of interest during the period of time.

The aspect of research efficiency should be considered with respect to both the academic output and the resources used to generate the output. There are multiple kinds of resources but the most essential ones are team members, research grants, and directly involved infrastructures. In other words, after a paper or a project is finished, we know how many members have contributed, how much money has been spent, how many groups have collaborated with the team of interest, and what facilities/equipment have been utilized. That is, the resource measure $R$ seems related to the number of coauthors, the number of grants, and the involved laboratories. Of course, the list of resources can be further enriched, and there are interactions among these items. To compare leaders in the same field or similar ones, we consider either the number of team members or the sum of research dollars as a good measure. Without loss of generality, we define the efficiency index $E$ as the ratio between $O$ and $R$.

$$E = \frac{O}{R} \tag{3}$$

We ought to treat the academic output and the research efficiency with equal importance. With other conditions being equal, the higher the output, the stronger the leadership; and the higher the efficiency, the stronger the leadership too. Doubling the output is as valuable as doubling the efficiency for a leader to be respected. Ideally speaking, given the same academic output, if the efficiency of a team is doubled, the saved resources can be used to produce more results, which is equivalent to double the academic output. On the other hand, at the same research efficiency, if the size of a team is doubled, the output of the team will be also doubled, which indicates that strong leadership has been demonstrated. Hence, to quantify the leadership, the output and the efficiency can be viewed as exchangeable in a multiplicative relationship. We argue that the output and the efficiency are quite independent. On one hand, one could produce a good amount of output but waste many resources. On the other hand, one might do just a small amount of work on something he/she is really good at with a high efficiency. To combine these two aspects into a single number, the product of $O$ and $E$ is preferred, instead of their sum. In other words, the leadership index $L$ should be a function of $O*E$.

$$L = f(OE) \tag{4}$$

We further argue that the geometric mean is suitable to convert the product $O*E$ into a linear scale (33). Indeed, the geometric mean gives a single figure of merit related to different factors in various ranges. For example, the geometric mean gives an average to compare two companies in terms of their environmental sustainability and financial viability. Many other geometric mean examples can be found as well. The use of a geometric mean normalizes the ranges of different factors so that all these factors are fairly treated. As a result, a given percentage change in any factor would induce the same effect on the overall measurement. Geometrically, the geometric mean of two variables gives the side length of the square whose area is the same as the rectangle defined by the two quantities. Hence, we propose that

$$L = \sqrt{OE} \tag{5}$$

Our main conceptual contribution is the leadership index in Eq. (5). Many leaders in history and at the present time are often described qualitatively, and augmented with a set of numerical evidence such as the awards he/she received, how many papers and books he/she wrote, how high his/her $h$-index is, and so on. A dedicated bibliometric measure of leadership has not been attempted before. We hope that our formulation will inspire further work along this line. Eq. (5) can be put in an equivalent form by inserting Eq. (3) into Eq. (5):

$$L = \frac{O}{\sqrt{R}} \tag{6}$$

Since Eq. (6) explicitly incorporates the resource factor into the leadership measurement, any leader should be aware of the negative impact when he/she requests too much funding. Within his/her capability, the more resources his/her team has, the higher the expected increase in academic output, which is more or less linear (the team is managed in the same efficiency with some degree of synergy and/or overhead). However, when the resources are overwhelming for him/her, the output increment rate will be reduced, and the *L* index will be compromised.

While the measurement of *O* is well known (see Eq. (1)), the measurement of *R* must be clarified so that the *L* index becomes computable. As a starting point, we underline that among all the resources, the human resource is the most essential. Also, we argue that the most essential leadership skill is to be able to identify, recruit, utilize, and keep talents to their full potential. Before the big data infrastructure becomes much more advanced, it is rather challenging to monitor the funding strength and flow, and even more challenging is to quantify the infrastructural factors. Hence, in this initial study we propose to use the following simplified yet insightful leadership index:

$$L = \frac{O}{\sqrt{T}} \tag{7}$$

where *T* is the total equivalent time/effort spent for the output *O*. As seen below, the beauty of Eq. (7) is that all the quantities are immediately accessible in current bibliometric databases, opening a wide window of opportunities to judge academic leaders bibliometrically.

Let us now explain how to compute *T*. Without loss of generality, the total output is here measured by the sum of *IF*s for a team of interest, assuming that all the papers are from a single team. Suppose that in a given period of time, the team leader has published *N* papers as the corresponding author with $IF_i$, $i=1, 2, ..., N$. Clearly,

$$O = \sum_{i=1}^{N} IF_i \tag{8}$$

In this period *P* (without loss of generality, let us assume P=1), the leader may have not spent his/her time equally on these papers. It would be reasonable to claim that he/she has spent more time on papers with higher *IF*s. Explicitly, his/her relative effort on the *j*th paper with $IF_j$ is estimated as $IF_j / \sum_{i=1}^{N} IF_i$. Furthermore, with the A-index we can compute the leader's credit share of the *j*th paper, which is equal to $A_j*IF_j$. The rest of the work has been performed by his/her team members. In other words, the amount of the leader's equivalent time spent on this paper is $\frac{IF_j}{A_j \cdot IF_j} IF_j / \sum_{i=1}^{N} IF_i$. Therefore, the equivalent time the leader has managed is computed as follows

$$T = \frac{1}{\sum_{i=1}^{N} IF_i} \sum_{j=1}^{N} \frac{IF_j}{A_j} \tag{9}$$

Note that the time equivalent argument is in the unit of the leader's time scale, which is the talent pool the leader directed in completing the paper. Various leaders have different time scales or efficiencies but in the same period of time the outputs of these teams are measured in the same unit, which is the sum of *IF*s. In other words, different leaders or team qualities will be reflected in terms of the common IFs and the expected number of citations, and calibrated against his/her equivalent time.

Then, we have the following leadership index that is immediately computable from the most popular bibliometric data:

$$L = \frac{\sum_{i=1}^{N} IF_i}{\sqrt{\frac{1}{\sum_{i=1}^{N} IF_i} \sum_{j=1}^{N} \frac{IF_j}{A_j}}} \quad (10)$$

which can be simplified to

$$L = \frac{\left(\sum_{i=1}^{N} IF_i\right)^{1\frac{1}{2}}}{\sqrt{\sum_{j=1}^{N} \frac{IF_j}{A_j}}} \quad (11)$$

*Toughness Index*

The above-formulated leadership index allows various measures of the scientific output or the value of a paper. While the number of papers is too rough, the number of citations is a function of time, and suffers from a major time lag. In this study, without loss of generality we focus on the weighted impact factor of a journal as a surrogate. The current bibliometric studies mostly rely on either the impact factor or the number of citations, without further treatment with respect to the magnitude of value. Here we argue that a high impact article such as one in Nature or Science cannot be equivalent to a number of ordinary specialty journal articles whose impact factor values are summed to be comparable to that of Nature or Science. In an extreme case, a collection of papers on classic mechanics cannot be equivalent to the single original paper on special relativity. Hence, we propose to introduce a toughness index on a logarithmic scale to place more emphasis on impactful original research.

Dr. Lev Davidovich Landau was a Soviet physicist contributing fundamentally to theoretical physics. He ranked physicists on a logarithmic scale from 0 to 5. Newton and Einstein were ranked 0 and 0.5 respectively. The fathers of quantum mechanics were ranked 1, including Bohr, Heisenberg, Dirac and Schrödinger. Physiologically, both our hearing and vision systems work on a logarithmic scale to handle wide dynamic ranges of audio and visual signals. Also, our information technologies are coded in binary sequences on a logarithmic scale. These observations motivate us to compute the toughness index as follows.

There are 49,205 journals in the period from 2010 to 2014 that were included for SCI analysis. Statistically, the number of citations to each journal divided by the journal impact factor gives the number of papers in that journal, resulting in 85,696,000 papers published in the five-year period. Then, we assign all the papers into ten levels so that a lower level has twice as many papers as an adjacent higher level. Suppose that X is the number of papers on the highest level, and Y is the total number of papers, we have $Y = \sum_{i=1}^{10} X^{i-1} = X(2^{10} - 1)$, and $X = \dfrac{Y}{2^{10} - 1} = 167,375$. Then, we sorted all the papers according to their impact factors from the highest to the lowest. The first set of 167,375 papers were assigned to the highest level; the second set of 2*167,375 papers went to the next highest level; the third set of 4*167,375 papers were for the third level; so on and so forth until the tenth level contained 2^9*167.375 papers. In this pilot study, we used 10, 9, …, 1 as weights for the ten levels form the highest to the lowest, respectively, which is illustrated in Figure 4. From now on, whenever we mention IFs, we mean weighted IFs and interpret the aforementioned equivalent time in this way.

*Data from National Natural Science Foundation of China*

In the first bibliometric study, we targeted life science research in China. We randomly selected 1,054 projects out of all 2,108 projects funded in 2007 by the National Natural Science Foundation of China (NSFC, http://www.nsfc.gov.cn). Each project has a unique identification number. The application codes are in the range from H01 to H31, representing 31 categories of medicine, including basic medicine, clinical medicine, public health, special medicine, and so on. The principal investigator's name, affiliation, budgetary information, project period, and other relevant information were retrieved from the database on the homepage of the National Natural Science Foundation of China. Moreover, reputable commercial websites were consulted, such as http://my.xywy.com, http://www.haodf.com, http://www.xuebang.com.cn, and the CNKI database of master theses and PhD dissertations, to double check and secure all the relevant information on age, gender, and rank. The data collection was performed from January to September, 2014. Table 4 summarizes the characteristics of the selected 703 principal investigators.

With the principal investigator's name and affiliation as the key for search, their respective journal papers were retrieved using the Web of Science for 2008-2013. Each and every paper was further verified for positive hits in terms of research direction and coauthors to exclude false alarms, in reference to the website of the National Natural Science Foundation of China. The resultant papers were archived into a master text file. This search was performed from July 2014 to May 2015. Then, the text file was transferred into an *excel* file, and the data were analyzed with *Visual Basic* and *SPSS 13*. The field values readable on the Web of Science were filled in for each of the retrieved papers, including the journal name and publication year.

The IF values were automatically assigned to the journals according to the data from the *Journal Citation Reports*. The number of papers was obtained for each principal investigator who served as the corresponding author for the papers.

Our selected principal investigators published 18,070 papers in total, among which 6,832 papers indicated them as the corresponding authors. The total number of corresponding authors is 783; that is, 266 of the 1,054 principal investigators had not published or cannot be confirmed as the corresponding author. Excluding 85 researchers not associated with any university in the period of 2008-2013 and limited to those whose papers were supported by the grants under evaluation, the selected 703 principal investigators collectively published 6,256 papers as corresponding authors. Furthermore, these principal investigators were stratified into three tiers. Among them, we placed 166 into the first class consisting of the top 10 universities in China, 358 of them into the second class from the rest of the top 100 universities in China, and the remainder (179) into the third class. The A-index value of the corresponding author was computed for each of his/her qualified papers, along with the impact factor and toughness index values.

*Data from National institutes of Health in United States*

In the second study, we retrieved the R01 projects funded in 2009 from the NIH official database (https://projectreporter.nih.gov/reporter.cfm), randomly sampled them at an approximate 1:10 ratio, and obtained 1,495 projects after redundancies were eliminated. The principal investigators were followed up from 2010 to 2014 to compute the total amount of funding to each principal investigator. Then, according to the NIH grant number, the papers published by each principal investigator were retrieved in the Web of Science database.

The data processing procedure similar to what we described in the preceding subsection was applied to the data collected for these NIH principal investigators. It was found that among the 1,495 R01 projects, only 832 of them resulted in original journal papers with the principal investigators as the corresponding authors, which put the total number of qualified papers at 3,034. Also, according to the 2016 US News Report Ranking of American Medical Schools, these NIH principal investigators were put into the first, second, and third classes corresponding to their affiliations in the top 10 (213 principal investigators), the rest of the top 50 (424), and outside the top 50 (195). Furthermore, the A-index value of the corresponding author was computed for each paper, along with its impact factor and toughness value.

*Experimental Design*

The two datasets from China and the USA are representative of the two major developed and developing countries. From Chinese and American datasets divided into three groups according to the most popular institutional ranking system in each country, systematic bibliometric analyses were performed to offer a quantitative view of Chinese and American academic capabilities with our proposed leadership index and its component variables, including academic output, research efficiency, group size (also measured as the principal investigator's equivalent time), and total funding. The leadership index was longitudinally analyzed to compare the trends of growth in China and the USA respectively.

**Conclusion**
In conclusion, for the first time we have proposed a bibliometric index of academic leadership. Based on an extensive survey on the peer-reviewed journal papers by representative teams of Chinese and American principal investigators, we have analyzed the leadership of these principal investigators keyed to the academic rank of their universities. Our methodology has potential to be extended to other bibliometric datasets to guide academic evaluation and resource allocation.

**Acknowledgments**
The authors are grateful for technical assistance by Dan Yang, Qian Yang, Anqi Wang, Faming Li, Yazhou Wang, Yu Wang, Yangyang Zhao, Li Wang, Lin Wang and Wenbo Liu with China Medical University.

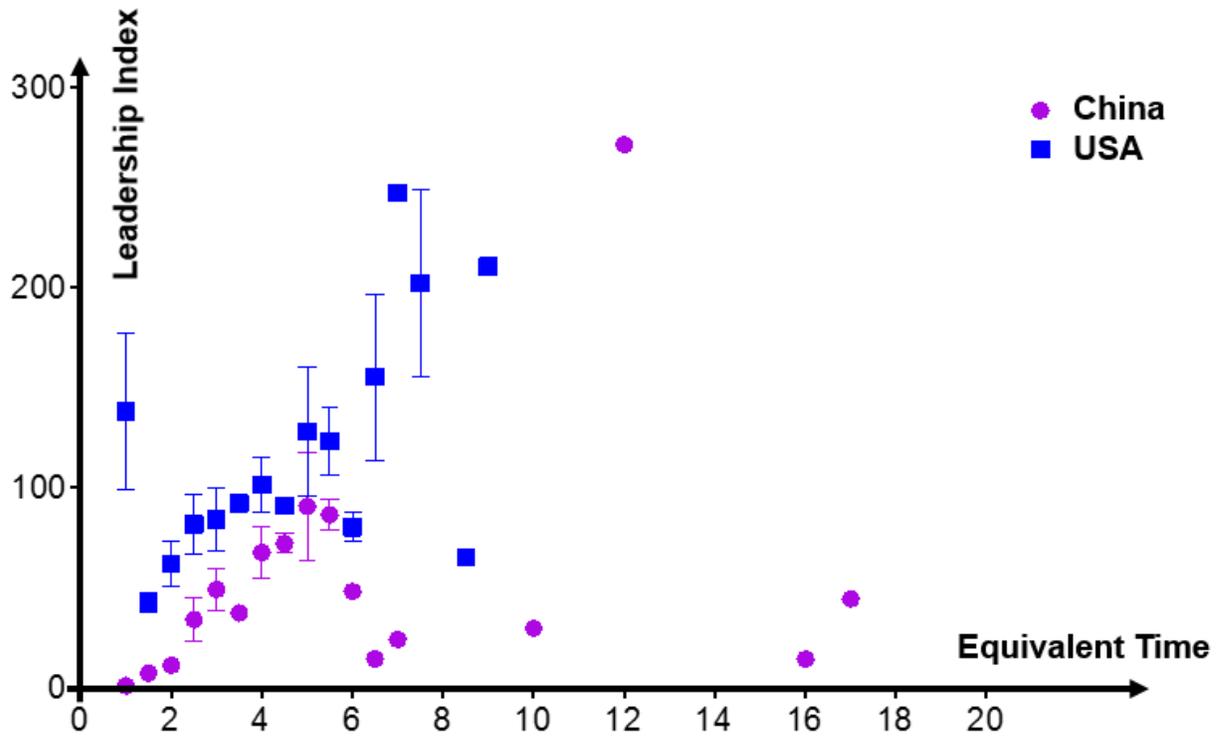

***Figure 1. Relationships between the academic leadership index and equivalent time in China and the USA.*** *The mean leadership indices are shown at discrete 0.5 intervals of equivalent time. Note that the pink circles and blue squares are for China and the USA, respectively.*

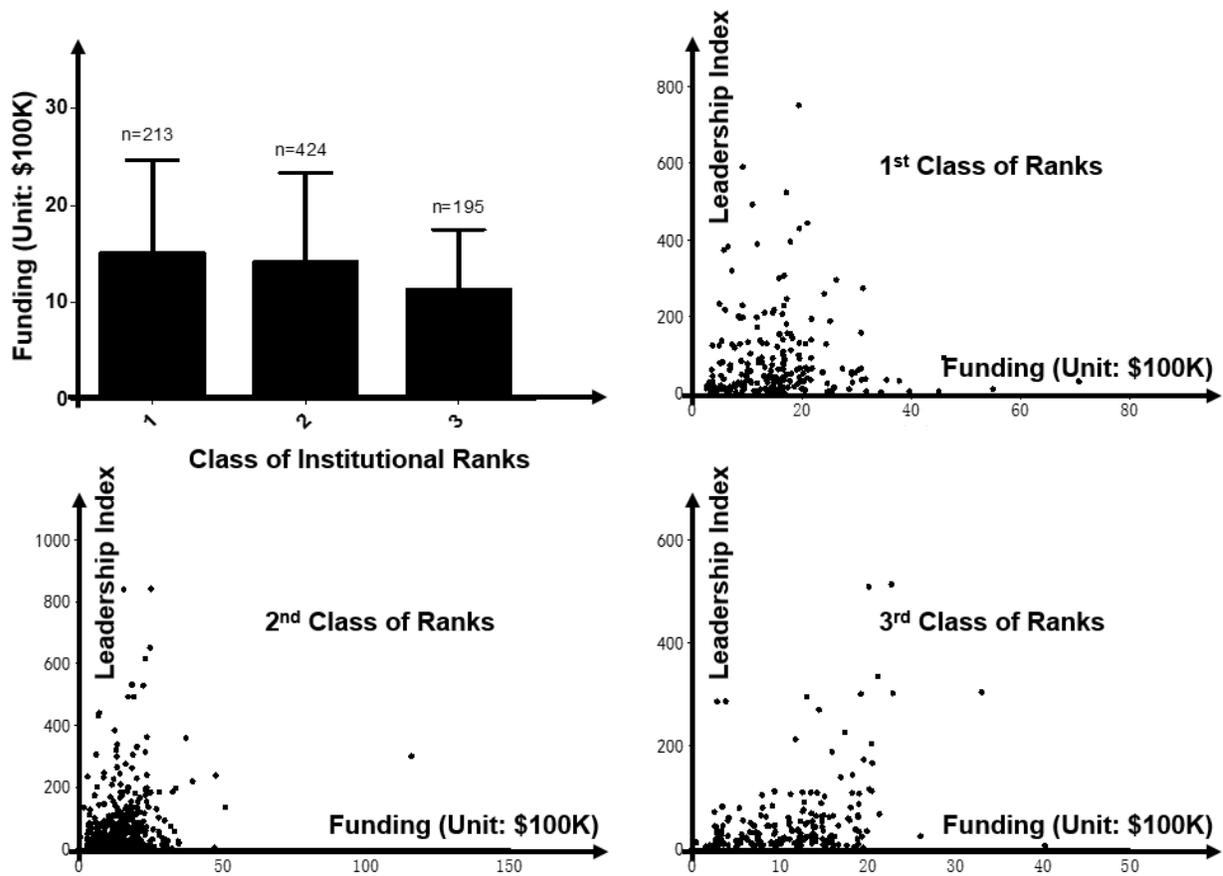

*Figure 2. Relationships between the leadership index and total funding in each class of universities.* The leadership index of principal investigators relative to their total funding is shown, keyed to the class of their institutes.

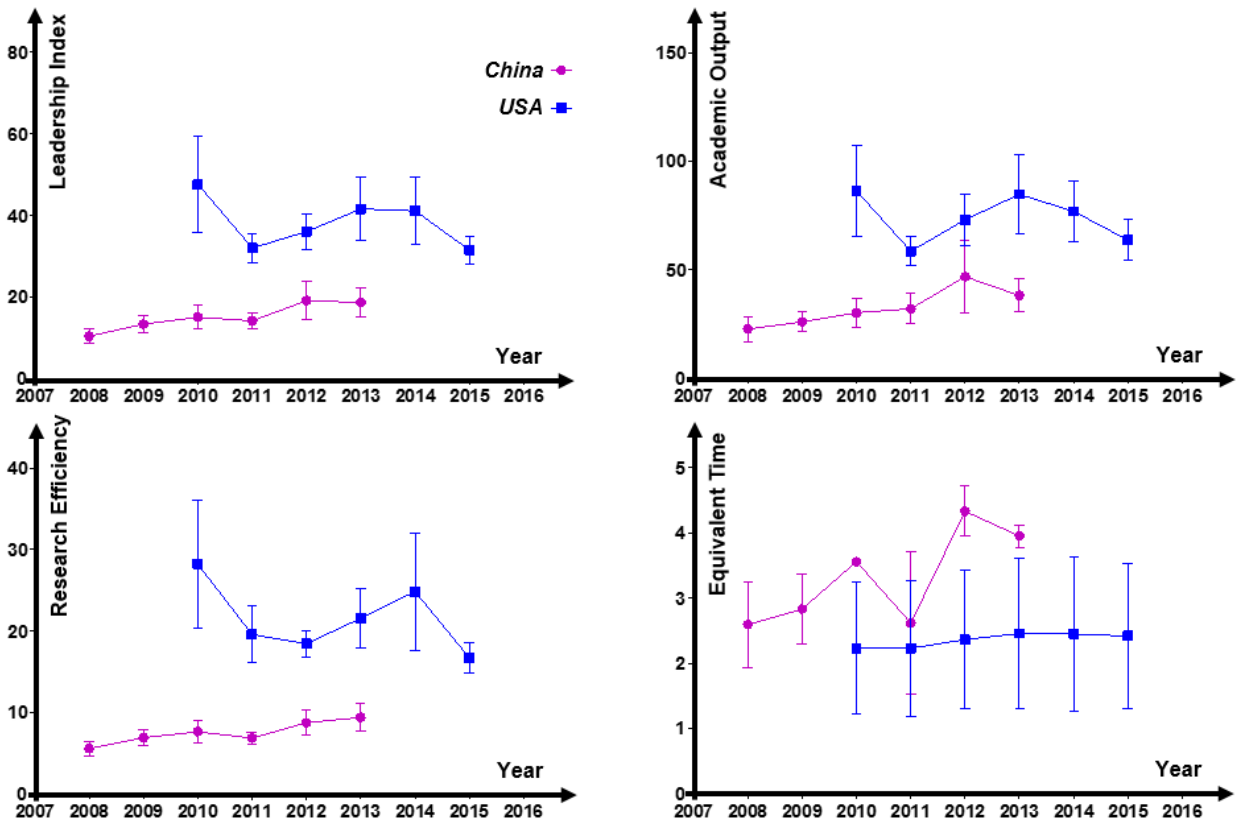

*Figure 3. Annual trends of academic leadership, scientific output, research efficiency, and equivalent time for Chinese and American principal investigators. The annual trends in each country were followed over a six-year span, beginning in 2008 for China and 2010 for the USA. Note that the pink circles and blue squares are for China and the USA, respectively.*

Table 1. Comparison of the leadership index values with respect to class and country.

| Index | University Class | China | | USA | |
|---|---|---|---|---|---|
| | | N | M±SD | N | M±SD |
| Academic Output without Weighting O' | 1 | 166 | 27.40±37.50 | 152 | 25.85±29.18 |
| | 2 | 357 | 21.18±29.77 | 365 | 24.94±27.96 |
| | 3 | 178 | 21.77±40.28 | 315 | 22.16±25.47# |
| | Total | 701 | 22.80±34.63 | 832 | 24.05±27.29 |
| Academic Output with Weighting O | 1 | 166 | 106.31±193.44 | 152 | 173.17±240.15## |
| | 2 | 358 | 72.03±109.84* | 365 | 148.90±210.87## |
| | 3 | 179 | 74.76±147.77* | 315 | 129.30±190.19*## |
| | Total | 703 | 80.82±143.77 | 832 | 145.91±209.45## |
| Equivalent Research Time | 1 | 166 | 3.92±2.90 | 152 | 3.43±1.27 |
| | 2 | 358 | 4.09±5.14 | 365 | 3.55±1.08 |
| | 3 | 179 | 3.42±1.06 | 315 | 3.55±1.08 |
| | Total | 703 | 3.88±3.97 | 832 | 3.53±1.11 |
| Research Efficiency | 1 | 166 | 24.79±35.80 | 152 | 52.46±76.60## |
| | 2 | 358 | 18.40±27.93* | 365 | 40.94±54.12*## |
| | 3 | 179 | 18.84±32.52 | 315 | 34.86±45.02**## |
| | Total | 703 | 20.02±31.19 | 832 | 40.74±56.15## |
| Leadership Index | 1 | 166 | 50.37±77.28 | 152 | 93.60±129.68## |
| | 2 | 358 | 36.18±54.88* | 365 | 77.16±103.22## |
| | 3 | 179 | 37.31±68.92 | 315 | 66.52±91.12**## |
| | Total | 703 | 39.82±64.61 | 832 | 76.14±104.65## |

Note that scientific outputs O and O' denote the sum of IF values with and without the involvement of the toughness index, respectively; i.e., the sum of IF values with and without weighting. For comparison between the first class and other classes, * for $P<0.05$ and ** for $P<0.01$. For comparison between China and the USA, # for $P<0.05$ and ## for $P<0.01$.

*Table 2. Chinese leadership index keyed to gender, natural age, and academic rank.*

|  |  | Number | Leadership Index |
|---|---|---|---|
| Gender | Male | 481 | 45.62±70.76 |
|  | Female | 194 | 29.14±49.03 |
| Age | Under 36 | 13 | 13.60±17.41 |
|  | 36-40 | 128 | 16.90±35.50 |
|  | 41-45 | 148 | 29.15±43.98 |
|  | 46-50 | 141 | 48.19±83.30 |
|  | 51-55 | 127 | 48.80±66.03 |
|  | 56-60 | 72 | 72.54±82.62 |
|  | Over 60 | 45 | 47.56±72.64 |
| Rank | Professor | 462 | 51.42±68.86 |
|  | Assoc. Professor | 182 | 19.87±53.93 |
|  | Assist. Professor | 27 | 5.29±5.86 |

*Table 3. Statistical testing results for the data in Table 3.*

| Test with Gender | | Male | | | Female | | | |
|---|---|---|---|---|---|---|---|---|
| | Male | / | | | 0 | | | |
| | Female | 0 | | | / | | | |
| Test with Age | | <36 | 36-40 | 41-45 | 46-50 | 51-55 | 56-60 | >60 |
| | <36 | / | 0.86 | 0.39 | 0.06 | 0.05 | 0 | 0.09 |
| | 36-40 | 0.86 | / | 0.11 | 0 | 0 | 0 | 0 |
| | 41-45 | 0.39 | 0.11 | / | 0.01 | 0.01 | 0 | 0.09 |
| | 46-50 | 0.06 | 0 | 0.01 | / | 0.94 | 0.01 | 0.95 |
| | 51-55 | 0.05 | 0 | 0.01 | 0.94 | / | 0.01 | 0.91 |
| | 56-60 | 0 | 0 | 0 | 0.01 | 0.01 | / | 0.04 |
| | >60 | 0.09 | 0 | 0.09 | 0.95 | 0.91 | 0.04 | / |
| Test with Rank | | Professor | | | Assoc. Prof. | | Assist. Prof. | |
| | Professor | / | | | 0 | | 0 | |
| | Assoc. Prof. | 0 | | | / | | 0 | |
| | Assist. Prof. | 0 | | | 0.26 | | / | |

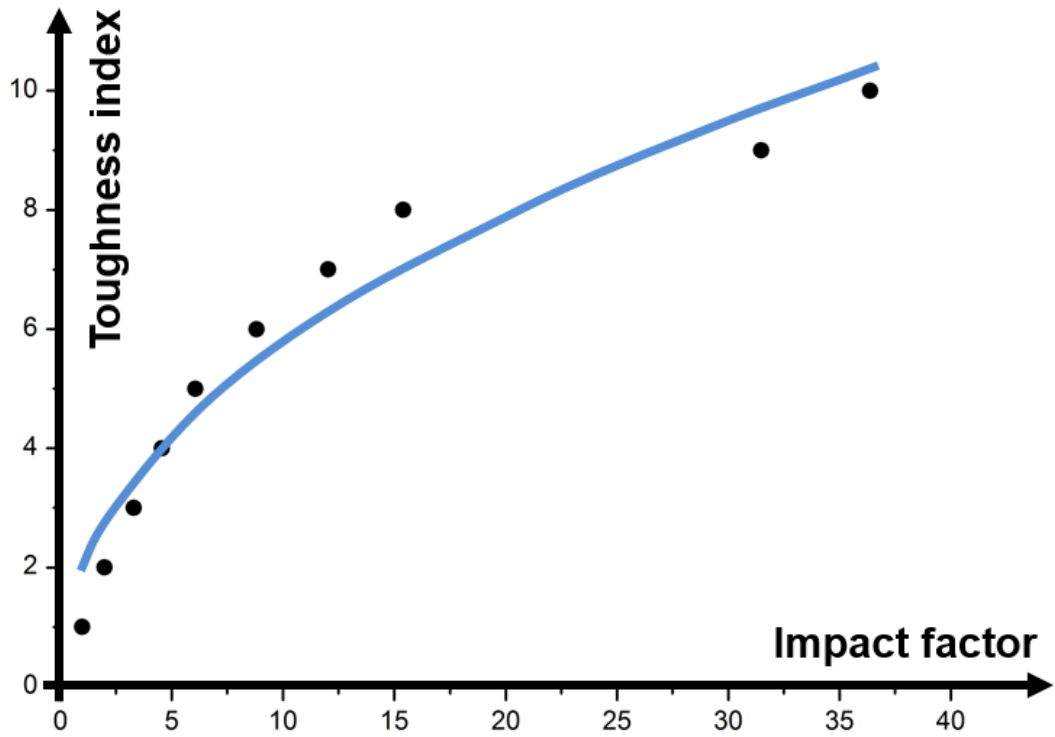

*Figure 4. Relationship between toughness index and journal impact factor.* Toughness index of a paper is defined in terms of the journal impact factor on a logarithmic scale, where the horizontal axis is for the impact factor and the vertical axis is for the toughness index.

*Table 4. Gender, age, and rank information of the selected 703 Chinese principal investigators.*

| Category | Detail | Number |
|---|---|---|
| Gender | Male | 481 |
| | Female | 194 |
| | Unclear | 28 |
| Age | Under 36 | 13 |
| | 36-40 | 128 |
| | 41-45 | 148 |
| | 46-50 | 141 |
| | 51-55 | 127 |
| | 56-60 | 72 |
| | Over 60 | 45 |
| | Unclear | 29 |
| Rank | Professor | 462 |
| | Associate Professor | 182 |
| | Assistant Professor | 27 |
| | Unclear | 32 |